\documentclass[preprint,preprintnumbers,amsmath,amssymb,elsart]{revtex4}
\usepackage{graphicx,psfrag,bbm,latexsym,color,dcolumn,bm,dsfont,bbm,color,mathrsfs,bbold,latexsym,amsmath,amsfonts,
amssymb,epsfig,natbib}

\newcommand{\beq}
{\begin{equation}}

\newcommand{\eeq}
{\end{equation}}

\newcommand{\beqa}
{\begin{eqnarray}}

\newcommand{\eeqa}
{\end{eqnarray}}

\newcommand{\bw}
{\begin{widetext}}

\newcommand{\ew}
{\end{widetext}}

\begin{document}

\title{The Partition Function Zeroes of Quantum Critical Points}
\author{P.R. Crompton}
\affiliation{Department of Applied Maths, School of Mathematics, Univerity of Leeds, Leeds, LS2 9JT, UK.}
\vspace{0.2in}
\date{\today}

\begin{abstract}
\vspace{0.2in}
{The Lee-Yang theorem for the zeroes of the partition function is not strictly applicable to quantum systems because the zeroes are defined in units of the fugacity $e^{h\Delta\tau}$, and the Euclidean-time lattice spacing $\Delta\tau$ can be divergent in the infrared (IR). We recently presented analytic arguments describing how a new space-Euclidean time zeroes expansion can be defined, which reproduces Lee and Yang's scaling but avoids the unresolved branch points associated with the breaking of nonlocal symmetries such as parity. We now present a first numerical analysis for this new zeros approach for a quantum spin chain system. We use our scheme to quantify the renormalization group flow of the physical lattice couplings to the IR fixed point of this system. We argue that the generic Finite-Size Scaling (FSS) function of our scheme is identically the entanglement entropy of the lattice partition function and, therefore, that we are able to directly extract the central charge, $c$, of the quantum spin chain system using conformal predictions for the scaling of the entanglement entropy.}

\vspace{0.1in}
\end{abstract}

\maketitle

\section{introduction}
The quantum spin chain system has been important for the development of much of the modern understanding of the role and nature of quantum fluctuations in low-dimensional quantum spin systems. The three important results for this system were derived in the 1980's and 1990's from treatments parameterised in terms of local fluctuations of the vacuum angle $\theta$, at finite values of $\theta$, where $\theta$ is defined as the prefactor of the topological term,

\beq
Q = -\frac{i}{2} \int \! dt \! \int \! dx \,\, \textbf{n}.(\partial_{x}\textbf{n} \times \partial_{t}\textbf{n})
\eeq

with $\textbf{n}$ a state vector for the spins in the chain, and $t$ and $x$ the imaginary-time and spatial coordinates of the model \cite{afm}\cite{the3}. Firstly, it was found that the infrared (IR) fixed point of the quantum spin chain system at $\theta=\pi$ is directly related to the unstable renormalization group flow of the SU(2) Wess-Zumino-Witten model, such that the central charge of the quantum spin chain system at $\theta=\pi$ is $c=1$ \cite{afm}. Secondly, it was identified in \cite{poly} that the quantum spin chain system is asymptotically free and, thirdly, it was established in \cite{fat} that the action of the quantum spin chain can be described effectively by the Sine-Gordon model in the vicinity of the IR fixed point at $\theta=\pi$. These three results indicate that conformal symmetries are important for classifying the quantum spin chain system, and also, that a nonperturbative description of the system would be needed in order to treat the quantum fluctuations of quantum spin chain systems at arbitrary values of $\theta$.

Formally, the procedure of renormalization defines how the scale of the fluctuations of a given system are related to the physical couplings. However, no procedure has yet been defined to renormalize the fluctuations in $\theta$ of the quantum spin chain at arbitrary $\theta$. Although it was established in \cite{fat} that perturbative renormalization is a well defined program for the effective action given by the Sine-Gordon model at the IR fixed point, it is not clear the extent to which this renormalization program can be generalised to arbitrary values of $\theta$. In fact, it has been argued more recently in \cite{mus} that the nonlocal fluctuations of $\theta$ are equally as important to the quantum spin chain system as the local fluctuations considered previously. Since the quantum spin chain is asymptotically free \cite{poly}, we have recently argued in \cite{me} that the full renormalization program for the quantum spin chain at arbitrary $\theta$ requires a nonperturbative renormalization scheme. The purpose of this article is to establish such a nonperturbative renormalization prescription for the quantum spin chain and, consequently, to quantify the link between the physical couplings of a quantum spin chain experimental system and the underlying conformal symmetries.

One way to treat the nonperturbative dynamics of the quantum spin chain is via a numerical lattice simulation using the Quantum Monte Carlo (QMC) method. This approach defines a realisation of the nonperturbative vacuum of the quantum spin chain that is given by evaluating the lattice partition function of the system via importance sampling. The approach
generates a finite statistical ensemble to represent the lattice partition function, which consists of a set of different configurations of the lattice generated at a fixed value of the inverse temperature, $\beta$, the interaction coupling between the spins, $J$, and the length of the quantum spin chain, $L$. There are two immediate problems with trying to define the nonperturbative renormalization group flow of the quantum spin chain via the QMC numerical procedure. Firstly, the quantum spin chain systems generated by the QMC method are restricted to finite values of $\beta$, whereas the IR fixed point of the system, identified in \cite{afm}, is at $\beta=\infty$. Secondly, it is not possible to give an area preserving definition for $\theta$ as a function of the lattice volume \cite{the3}\cite{the1}\cite{the2} and, consequently, to define how $\theta$ fluctuates as a function of lattice system size. However, we have recently resolved both of these issues algebraically in \cite{me}, \cite{me4} and \cite{me7} by defining a new Wick rotation procedure for the local transfer matrix elements of the QMC method, in order to define an exact mapping between $\beta$ and $\theta$. Although there is no critical point in the quantum spin chain system at finite temperature, we are able to define the line of constant physics in the $\beta -J$ plane that flows into the quantum critical point from this scheme.

Our new Wick rotation procedure is essentially a generalisation of the properties of partition function zeros, as we
discussed in detail in \cite{me4}. The reason for this is that both of these schemes are based on a mathematical procedure called meromorphic continuation. Partition function zeros are conventionally defined through an exact polynomial expansion of the lattice partition function defined over the local degrees of freedom of a lattice system \cite{l+y2}, and the zeroes of the finite polynomial expansion defined for a finite volume lattice system uniquely define the singularities of the lattice partition function of that finite system. However, there is also a nontrivial relationship between the zeroes of a finite lattice system and the zeroes of that same system treated in the infinite lattice volume limit \cite{l+y2}. This relationship is defined formally by a meromorphic mapping.

To define our new Wick rotation procedure, we have used the formal definition of a meromorphic mapping to treat a more
generalised polynomial expansion than the conventional local polynomial expansion, which includes the nonlocal degrees of freedom of the lattice model of a quantum spin system in addition to the local degrees of freedom. As we have rigorously proven in \cite{me4}, we can, therefore, now use this new polynomial expansion to quantify the singularities of the partition function of a lattice system associated with the nonlocal fluctuations of $\theta$, as well as the local fluctuations of $\theta$. This determination defines our new nonperturbative renormalization program for fluctuations in $\theta$, since it completely defines the fixed points of the nonperturbative vacuum in $\theta$ in the mapping to the infinite lattice volume limit. The numerical analysis we present is, therefore, simply a repetition of the conventional Finite-Size Scaling (FSS) analyses given in \cite{z1}-\cite{zn} for the conventional partition function zeros of classical spin systems, but now applied to our new form of nonlocal polynomial expansion. The only significant difference that arises between this new zeros analysis and the analyses in \cite{z1}-\cite{zn} is that the new generic FSS function of the zeros has the form of the entanglement entropy of the system, as we discuss in Section 7. This means that we are then able to compare the results of our new zeros analysis with conformal results on the FSS of the entanglement entropy of the quantum spin chain, and from this we deduce the central charge of our lattice ensembles.

One practical point remains, which is that although our Wick rotation procedure defines the meromorphic continuation
properties of $\theta$ on the lattice, we would still like to be able to vary the fixed value of $\theta$ at which our
lattice ensembles are generated in order to generate lattice ensembles at different values of $\theta$. The reason for this is that this will allow us to assess the sensitivity of our new renormalization approach to perturbations in $\theta$. Although it would be possible for us to generate lattice ensembles directly at the point $\theta=\pi$ by simulating models consisting entirely of half-integer spins with the same spin magnitudes, it would not be possible to vary $\theta$ by varying any of the lattice coupling parameters in these models. We, therefore, choose to simulate the model which consists of an antiferromagnetic (AFM) periodic mixed-spin chain of the form, $1-1-3/2-3/2$, where $1$ and $3/2$ represent the spin magnitudes, and the period of the alternation is four lattice sites \cite{me2}\cite{me3}. We can vary the fixed value of $\theta$ at which a lattice ensemble is generated in this model by varying the ratio of the nearest-neighbour spin interaction couplings within the periodic cell. The Hamiltonian of this model is given by,

\beq
H = \sum_{j=0}^{L/4-1} J_{1 \,, 1} {\bm S}^{(1)}_{4j} . {\bm S}^{(1)}_{4j+1} +
J_{1 \,, 3/2} {\bm S}^{(1)}_{4j+1} . {\bm S}^{(3/2)}_{4j+2}  +
J_{3/2 \, , 3/2} {\bm S}^{(3/2)}_{4j+2} . {\bm S}^{(3/2)}_{4j+3} +
J_{3/2 \,, 1} {\bm S}^{(3/2)}_{4j+3} . {\bm S}^{(1)}_{4j+4}
\eeq

where $j$ is the lattice site index of the cell along the spin chain, $J_{a, b}$ is the nearest neighbour spin interaction coupling between the neighbouring spins of magnitude $a$ and $b$, and ${\bm S}^{(a)}$ is the spin operator of magnitude $a$. For our numerical simulations of the model we keep the ratio between like-like and like-dislike spin vectors fixed such that $J_{1 \,, 1} = J_{3/2 \,, 3/2} = J$ and $J_{1 \,, 3/2} = J_{3/2 \,, 1} = \alpha J$. We organise the article as follows. In Section 2 we review the definition of the transfer matrix of the continuous-time QMC method, and in Section 3 define how this can be used to give an exact polynomial expansion for the lattice partition function of the quantum spin chain. In Section 4 we discuss the symmetry properties of our new polynomial expansion, and how the zeros of the polynomial relate to the zero temperature fixed points of the quantum spin chain through Wick rotation. In Section 5 we identify the generic FSS behaviour of our new zeros density function, which we identify as the entropy of the QMC method. In Section 6 we present the numerical FSS analysis of an AFM periodic mixed-spin model of the form $1-1-3/2-3/2$. Finally, in Section 8, we identify that the entropy that we have measured is identically the entanglement entropy of the system, which follows from the bipartite definition of the partition function obtained from Trotter-Suzuki decomposition, and we compare our numerical FSS result with conformal predictions for the scaling of the entanglement entropy at the $\theta=\pi$ fixed point of the quantum spin chain system.

\section{the Transfer matrix}

\subsection{The Loop-Cluster Lattice Partition Function}
There are several closely related loop-cluster QMC methods for the numerical study of low-dimensional lattice quantum spin systems, and a comprehensive review of all of these schemes is given in \cite{rev}. Trotter-Suzuki decomposition is used to define the generic partition function for a number of these QMC methods, and in particular, the method we have used to generate our numerical data in this article. In this decomposition the spin operators, $S$, of a $D$ dimensional lattice quantum spin system are placed on a $D+1$ dimensional space-Euclidean time lattice as classical spins. The spin Hamiltonian, ${\cal {H}} = J \sum_{i} S_i S_{i+1}$, for a system defined with an AFM nearest-neighbour interaction is then split into a sublattice definition on the space-Euclidean time lattice, where the spin site index $i$ of the Hamiltonian, $H_i=J S_i S_{i+1}$, is defined only on the even-indexed spin sites, with $J$ is the AFM nearest-neighbour spin interaction coupling. Including different spin vectors does little to change the basic form of this decomposition, since higher order spins can be implemented locally through the inclusion of a permutation factor \cite{lu}. The lattice partition function of the generic $D$ dimensional quantum spin system with an AFM nearest-neighbour spin interaction is defined via Trotter-Suzuki decomposition through the following trace,

\beq
\label{Trotter}
e^{\beta \cal {H}} = \prod_{i=1}^{L}\exp(-\Delta\tau H_i + {\cal O}(\Delta\tau^{2}) \, )
\eeq

where $\beta$ is the inverse temperature, $\Delta\tau$ is the Euclidean-time lattice spacing, $i$ is the lattice site index, and $L$ is the number of sites on the lattice \cite{rev}. This recovers the quantum spin operators of the $D$-dimensional model from the classical spins of the $D+1$-dimensional model. For our analysis of the quantum spin chain $D=1$.
This choice of sublattice Hamiltonian used for the Trotter-Suzuki decomposition of a quantum spin system with an AFM
nearest-neighbour spin interaction means that the lattice partition function can be expressed via a basis of
two-by-two site blocks known as plaquettes \cite{qmc},

\beq
\label{matrix}
{\cal{Z}} = {\rm Tr} \, e^{\beta \cal{H}} = \prod_{i,t = 1,1 }^{L,T}
\langle S_{i,t  } S_{i+1,t  } | e^{-\, \Delta\tau \cal{H} } | S_{i,t+1} S_{i+1,t+1} \rangle \;
\eeq

where $\langle . \rangle$ denotes the transfer matrix for one of the elementary plaquettes defined in $i\otimes t$, where $i$ is the spatial lattice site index (in the $D$ direction), $t$ is the index of Euclidean-time sites on the lattice (in the $+1$ direction), and $T$ is the total number of sites in the Euclidean time ($+1$) direction \cite{rev}.

The transfer matrix $\langle . \rangle$ of an O(3) spin system with a nearest neighbour exchange is of the form,

\beq
\label{transfer}
\langle . \rangle = \left(
\begin{array}{cccc}
1 & 0 & 0 & 0\\
0 & \frac{1}{2} ( 1 + e^{-J\Delta\tau}) & \frac{1}{2} ( 1 - e^{-J\Delta\tau})  & 0 \\
0 & \frac{1}{2} ( 1 - e^{-J\Delta\tau}) & \frac{1}{2} ( 1 + e^{-J\Delta\tau}) & 0 \\
0 & 0 & 0 & 1 \end{array}
\right)
\eeq

In this definition the diagonal elements describe the possible Euclidean-time evolution of a spin defined on the spatial sites, and the off-diagonal elements describe what must subsequently happen to the spins on the nearest-neighbour spatial sites in order to locally conserve total spin via an antiferromagnetic exchange. The form of the transfer matrix that is given in (\ref{transfer}) is essentially the same as that used for the Density Matrix Renormalization Group, but the important difference for the QMC method is the way in which the above Trotter-Suzuki formalism can be used to define loop-clusters for the Monte Carlo procedure. The general approach of the QMC method is closely related to classical loop-cluster methods, as is discussed in detail in \cite{rev}.
\subsection{Continuous-time}
The general aim of applying the Monte Carlo scheme to lattice models is to generate a sequence of lattice configurations via a dynamical Markov process. An importance sampling decision separates each of the lattice configurations in this sequence (or Markov chain), and the set of all of these lattice configurations then defines a statistically meaningful ensemble and the lattice partition. The important consequence of this generic Monte Carlo approach for the evaluation of (\ref{matrix}) is that there are then in fact three discrete indices for the lattice partition function evaluated via the generic Monte Carlo method. The first of these indices is the spatial index, $i$, the second is the Euclidean-time index, $t$, and the third (which is usually dropped from notation) is the configuration index, $M$, which indicates the relative position of a lattice configuration within the Markov chain.

New lattice configurations are generated for the importance sampling step of the Monte Carlo process by making small random changes to the state of the spins on the lattice, and in the classical loop-cluster scheme these changes are made by changing the state of the spins on the lattice within a closed loop. The point of using this loop-cluster definition for the importance sampling of classical spin systems, as was treated rigorously in \cite{loop1}\cite{loop2}, is that the sum over discrete steps in the probabilistic Markov chain, $M$, can then essentially be interchanged with the sum over discrete loop-clusters in the definition of the lattice partition function, which leads to more efficient numerical sampling. As was noticed in \cite{qmc} and \cite{rev}, however, when this loop-cluster definition of the new lattice configurations is applied to the Trotter-Suzuki formalism in (\ref{matrix}), this definition has the advantage that closed loops on the $D+1$ dimensional lattice in (\ref{matrix}) automatically preserve the definition of the trace. Therefore, as is used explicitly in \cite{qmc}, using this approach it is possible to interchange, $t$, with, $M$, and to define the lattice partition via the with respect to a trace operation defined over the length of the Markov chain, rather than with respect to Euclidean-time. A formal definition of the continuous-time properties of the general continuous-time Monte Carlo method is given in \cite{math}, and it is explained in detail how, $t$, and, $M$, can be interchanged in a general Monte Carlo method by using continuous-time.

The QMC method we employ for our zeroes procedure is known as the continuous-time QMC method, and it corresponds to an
algorithm where, $t$, is interchanged with, $M$, in the definition of (\ref{matrix}). We define the lattice partition function of the continuous-time QMC method from \cite{qmc} by the following product of local transfer matrices,

\beq
\label{part1}
{\cal{Z}} = {\rm Tr} \, e^{\beta \cal{H}} = \prod_{i=1}^{L}
\left( \begin{array}{cccc}
1 & 0 & 0 & 0\\
0 & \lambda_i  & 1 - \lambda_i & 0 \\
0 & 1 - \lambda_i & \lambda_i & 0 \\
0 & 0 & 0 & 1 \end{array}
\right)
\eeq

where $i$ is the spatial lattice site index, and $\lambda_i$ are the local transition probabilities corresponding to the entries of the transfer matrix. To be clear, these transition probabilities, $\lambda_i$, are taken from the final figure in \cite{qmc}, but there is no explicit expression in \cite{qmc} that defines the lattice partition function
as a product of these matrix elements. Therefore, to obtain (\ref{part1}) we have substituted these transition probabilities from the final figure into the transfer matrix definition in (\ref{matrix}) (which is also from \cite{qmc}). Our reason for using a more general form of the transfer matrix than (\ref{transfer}) is that we want to distinguish the transfer matrix elements that are realised numerically for each lattice site, $i$, since these values are explicitly evaluated at each step of the Markov process and are used to build up a realistic probability distribution for the spins.

The most obvious difference from the way the above lattice partition function is written from (\ref{matrix}), however, is that the trace over the spatial and Euclidean-time directions of the $D+1$ dimensional lattice in (\ref{matrix}) have been replaced in (\ref{part1}) by a trace over just the spatial index, $i$. In fact, to write (\ref{part1}), the trace over $t=\{1,T\}$ has been replaced in (\ref{part1}) by a trace over just one site, $M=\{1,2\}$, through the exchange $t\leftrightarrow M$. To write the lattice partition function in this form we have used the continuous-time properties of the general continuous-time Monte Carlo method defined in \cite{math}. We have noticed, firstly, that both the number of discrete sites in Euclidean-time, $T$, and the number of discrete steps in the Markov chain, $M$, are arbitrarily defined for the partition function in (\ref{matrix}). Secondly, for a Markov process, the importance sampling process represents a probabilistic decision that is continuously defined in the interval $(0,1]$. Therefore, because of the more rigorous functional definition of the probabilistic interval that the steps in the Markov chain represent, there is no restriction on the number of steps in the Markov chain, and it is possible to represent the Markov chain by only one step. This is also the rigorous basis of dropping the usual statistical configuration index on lattice partition functions evaluated via the Monte Carlo method, \cite{math}.  To be clear, this is not the usual way that the boundary conditions of the continuous-time QMC method in \cite{qmc} are presented in review articles \cite{rev1}\cite{rev2}. However, the usual tacit assumption that is presented: that $M$ commutes with $t$, is only strictly true in the large volume limit. In a rigorous sense, the continuous-time method is a nonequilibrium Monte Carlo method \cite{math}, as we have presented it here, where the noncommutativity of $M$ and $t$ can be realised in the numerical simulation of a finite system.

The value of expression (\ref{part1}) for evaluating our new scheme is twofold. Firstly, it is straightforward to define an exact polynomial expansion for the lattice partition function in (\ref{part1}) (defined over $i$), as we will now show. Secondly, the $\{\lambda_i\}$ are defined to be a continuous functions over $(0,1]$ in the continuous-time method \cite{qmc}\cite{math}, which allows us to define our exact Wick rotation procedure for the lattice system.

\section{lattice partition function polynomial expansion}

To form a polynomial expansion over the spatial site index $i$ for the lattice partition function in (\ref{part1}), we first introduce, $P$, which is a $4i\times 4i$ block diagonal matrix, where the blocks are the local transfer matrices, $\langle . \rangle$, that define the matrices in the product in (\ref{part1}) and all other entries in $P$ are zero,

\beq
P = \label{transfer2}
\left( \begin{array}{cccc}
\langle . \rangle_{k,M} & 0                           & ...    & 0      \\
0                         & \langle . \rangle_{k+1,M} &        & \vdots \\
\vdots                    &                             & \ddots &     0  \\
0                         & ...                         & 0      & \langle . \rangle_{L,M} \\
\end{array} \right)
\eeq

An exact polynomial expansion for the lattice partition function, in $\beta$, is given by finding the characteristic equation of the eigenvalue problem ${\bf{det}} (P - \beta) =0$.

There is a precedent for using numerical lattice simulation data in this way to calculate the zeros of the lattice partition function \cite{me5}\cite{me6}. There are difficulties for practical implementing these schemes because the evaluation of the characteristic polynomial coefficients, $c_i$, from numerical input from standard numerical recursion can be subject to roundoff errors. This a general problem for
all exact diagonalization and Density Matrix Renormalization Group analysis that involve diagonalization, and often this
roundoff forms the dominant systematic error, as is discussed at length in \cite{ed1}\cite{ed2}\cite{ed3}. This source of
roundoff error is both a function of the number of sites in a lattice model and also the condition number of the system that
is numerically realised. Often, this roundoff restricts the range of lattice volumes that can be treated via an approach
involving diagonalization.
To address this problem in our analysis, we have implemented the tested techniques in \cite{me5} designed to minimise
roundoff. These techniques involve using Newton's relations to define the operation $c_i \equiv c_{i}[\lambda_{i}]$, in
combination with a series of logarithmic modifications to the standard numerical operations. We refer the interested reader
to these studies \cite{me5}\cite{me6} (and references therein) for further details of the numerical stability of these
techniques, but comment that the size of the coefficients that we have evaluated for the quantum spin chain system mean that
the roundoff problems from the routines themselves are negligible as a function of lattice volume since they have been
used to treat successfully treat polynomials of ${\mathcal{O}}(2500)$.

The application of Newton's relations is a standard linear algebra procedure for calculating characteristic polynomial
coefficients. The procedure consists of evaluating $c_i \equiv c_{i}[\lambda_{i}]$ recursively from powers of the trace of
$P$, as defined in \cite{me5}. We introduce our Wick rotation at this point in the analysis, mapping
$\beta \rightarrow i\beta$ through the following expression for the trace of the local transfer matrices in (\ref{part1}),

\beq
\label{trace}
{\rm Tr} \, \langle . \rangle = \rm{det}\! \left(
\begin{array}{cc}
1                  & \lambda_{i} \\
\lambda_{i} & 1     \\
\end{array} \right)
\eeq

This Wick rotation definition, following \cite{me}, \cite{me4} and \cite{me7}, changes the boundary conditions of the
lattice partition function from $(0,\beta] \rightarrow [-\pi,\pi]$. The trace of $P$ is then easily calculated from the product
of these local transfer matrix traces via,

\beq
{\rm Tr} P = \sum_{i=1}^{L}  ( 1 -\lambda_{i}^2 )
\eeq

The characteristic polynomial equation can then evaluated via Newton's method to find the polynomial coefficients in the
Wick rotated basis, and the lattice partition function polynomial expansion can be written as,

\beq
\label{poly}
{\cal{Z}} = \sum_{i=1}^{L+1} \, \langle c_{i} \rangle \, \beta^{2i}
\quad\quad : \quad\quad n\, c_{n} + {\rm Tr} P^{n} + \sum_{k=1}^{n-1} c_{k} {\rm Tr} P^{n-k} = 0
\eeq

where $i$ the spatial lattice site index, $\beta$ is the inverse temperature, $c_i$ are the polynomial expansion
coefficients, $n=1, ... , L$, and $k$ is a dummy index.

To summarise the steps we have taken to define this exact polynomial expansion for the lattice partition function of the
quantum spin chain: we have first taken the definition of the continuous-time lattice partition function in \cite{qmc} and
formulated this expression (\ref{part1}) as eigenvalue problem in $\beta$. We have then Wick rotated the local transfer
matrix elements following the algebraic procedure we have defined in \cite{me}, \cite{me4} and \cite{me7}, by defining the
trace operation in (\ref{trace}), to map $\beta \rightarrow i\beta$ and re-define the boundary values of the lattice
partition function from Euclidean to imaginary time. Finally, we have then proposed evaluating the characteristic equation of this eigenvalue problem using
Newton's relations and the numerical recursion modifications defined in \cite{me5} and \cite{me6}. This will then give a
semi-analytic exact polynomial expansion, defined over the lattice spatial site index $i$, which can be evaluated from
numerical input in the form of the local transition probabilities $\lambda_{i}$ realised in the importance sampling of the
continuous-time simulation of the quantum spin chain system.

\section{properties of the polynomial Zeroes}
In the previous section, we have described our analytic procedure for converting the numerical transfer matrix entries of
the continuous-time QMC method into an exact semi-analytic polynomial expansion for the lattice partition function of a
quantum spin system. We claim that the zeros of the lattice partition function in (\ref{poly}) should share similar scaling
properties to the partition function zeroes defined in \cite{z1}-\cite{zn} because we have used the algebraic Wick rotation
procedure we have defined in \cite{me}, \cite{me4} and \cite{me7} to define the polynomial.

One of the most important properties of the zeroes defined in \cite{z1}-\cite{zn} comes from an argument presented in the
Appendix of \cite{z1}. This states that if the local elements used to define the characteristic polynomial coefficients are
bounded in magnitude, then the magnitude of the coefficients is necessarily bounded, as is the magnitude of the zeroes. This
result is important to the FSS analysis of partition function zeroes because it means that the zeroes are then constrained
to lie on a locus in the complex expansion parameter plane, which simplifies the definition of the FSS function \cite{z1}.
The generic FSS function of the zeroes is given formally in the next section. However, it can be noted that for our new
expansion in (\ref{poly}) a similar bound exists because our polynomial coefficients are defined from the local transition
probabilities $\lambda_i$, which are bounded in the interval $(0,1]$.

For our Wick rotated procedure, the meaning of this locus constraint has a special connection with the way that O(3) spin
operators of the quantum spin chain are represented in the continuous-time method. Changing the continuous-time
representation of O(3) spins via the Wick rotation procedure in (\ref{trace}) means that the $S_z$-component of the O(3)
spin vectors is represented by the spin vectors defined on each spatial lattice site $i$, and the Euclidean-time interval,
$(0,\beta]$, represents the projection of the spins onto the plane of the $S_x$-$S_y$ components. Therefore, the lattice
nearest-neighbour spin interaction coupling $J$ only couples to the $S_z$-component of spin of the new Wick rotated operators,
and the coupling between the $S_x$ and $S_y$ components is implicitly normalised to unity. The bounds
that define the locus of our new expansion are, therefore, set by the physical surface traced out by the spin vectors. In an
isotropic system, where the interaction couplings to $S_x$, $S_y$ and $S_z$ are all unity, we can deduce that the zeros of
(\ref{poly}) should lie on a sphere because the spin vectors have a $4\pi$-rotational symmetry, ie. the locus is a sphere.
However, any deviations of $J$ from unity will break the isotropy of this surface traced out by the spin vectors. We can,
therefore, include the analytic factor $J$ into the exact polynomial in (\ref{poly}) in order to normalise the units of the
surface for our analysis,

\beq
\label{poly2}
{\cal{Z}} = \sum_{i=1}^{L+1} \, \langle c_{i} \rangle \, (\beta J)^{2i} \,.
\eeq

This then allows us to relate the zeroes locus to the lattice volume (defined in physical units), and to make the connection
between the critical scaling of the lattice partition function in $\theta$ and the physical units of the lattice ensemble.

\section{the Zeroes density}

The most important feature we have claimed for our expansion in (\ref{poly}) is that it shares the same FSS properties as
the partition function zeroes analyses defined in \cite{z1}-\cite{zn}. We, therefore, first review the FSS function
arguments for these classical spin systems, and then present a new FSS definition for our new expansion that we show is
equivalent.

\subsection{The Zeroes Density of the Ising Model}
The FSS function of the generic partition function zeroes analysis is defined by the density of the partition zeros along
the locus that they form in the complex plane of the expansion parameter \cite{z1}-\cite{zn}. For the Ising model system
considered in \cite{z1}, the zeroes density, $\lambda(\phi,t)$, is defined through a saddlepoint equation for the free
energy, $F$, of the following form

\beq
\label{density}
F = -\beta h \,\, -\!\!\int_{-\pi}^{\pi} \lambda(\phi,t) \,\, {\rm{log}}[e^h -e^{i\phi}]\,{\rm{d}}\phi
\eeq

where, $h$, is the applied external field in the Ising model, $t$ is the reduced temperature, $\phi$ is the angle that a
zero on the locus subtends to the real $e^h$-axis, and the expansion parameter for the partition function zeroes treatment
is the fugacity variable $e^h$. The zeros density, $\lambda(\phi,t)$, is also subject to the boundary conditions,

\beq
\label{normal}
\lambda(\phi,t)=\lambda(-\phi,t)\,\,;\quad\quad \int_{0}^{\pi}
\lambda(\phi,t) \,{\rm{d}}\phi = \frac{1}{2}
\eeq

The only stable solution for the free energy in (\ref{density}) occurs at $\phi=0$ and this indicates that the critical points of the
Ising model considered in \cite{l+y2} must lie on the real $e^h$-axis. This critical point is the saddlepoint solution for
the free energy, found in the asymptotic limit where the zeroes density becomes a continuous function. What is expected on a
finite lattice is that none of the zeroes will lie on the real $e^h$-axis, but that the zeroes density is still an
asymptotically convergent function towards the critical point. This assertion is used to argue that the zeroes density
function is a valid FSS function in \cite{l+y2}. Note, however, that there is no rigorous statement in this definition that proves that (\ref{density}) is the free energy of a lattice partition function. The statement to justify the partition function zeros method given in \cite{l+y2} is simply that the thermodynamic properties of a lattice system can be treated exactly if the free energy of a lattice partition function is written in the form of (\ref{density}).

\subsection{The Zeroes Density of the Quantum Spin model}
Unfortunately, we cannot make a simple substitution to map our new expansion variables into (\ref{density}) in order to define the zeroes density of our new formulation. The reason for this is that the boundary conditions of our new formulation are branch points, because of the way we have defined the Wick rotation which maps $(0,\beta] \rightarrow [-\pi,\pi]$. We start by noticing that the $\lambda_{i}$ in (6) have an explicit dependence on the couplings $\beta$ and $J$. This dependence arises through the bare parameter dependencies of the plaquette definition in (\ref{part1}), which are used to construct the $\lambda_{i}$ through importance sampling \cite{loop1}\cite{rev1}. Each of these numerically-realised local transfer matrix entries represents a transition probability, therefore, for these transition probabilities to be well-defined it follows that the integral over all values of the bare parameters must be one. This motivates a normalizing condition, analogous to (\ref{normal}),

\beq
\label{norm}
\lambda(\beta,J) \geq 0; \quad\quad \int_{0}^{\infty} \lambda(\beta,J)
\,{\rm{d}}J = 1 .
\eeq

In our notation the $\lambda_{i}$ are the local transfer matrix elements of a finite, discrete lattice system, whereas the function $\lambda(\beta,J)$ is the transition probability corresponding to these matrix elements defined for a continuous system, as a function of the continuous variable $J$. Note that the $\beta$-dependence of the local transfer matrix elements $\lambda_{i}$ is already made continuous by the definition of the continuous-time method in \cite{math}\cite{rev1}.

A rigorous treatment of the existence of this continuous-$J$ limit of the transition probabilities is given by applying the method of $\zeta$-function renormalization \cite{SH}, which has been developed for this lattice model in \cite{me4}. We define an analogous saddlepoint equation to the one in (\ref{density}) for the polynomial expansion in (\ref{poly}), by defining the analytic properties of the support of this lattice model. We do this via quantifying the logarithm of the determinant of the operator $P$ in (\ref{transfer2}), where,

\beq
{\rm {det}} P = \prod_{i=1}^{L} \Lambda_{i}.
\eeq

and $\Lambda_{i}$ are the eigenvalues of $P$, $\Lambda_{i}=2\lambda_{i}-1$.
To quantify this logarithm (whilst resolving any undetermined branch points in the continuous-system limit, $L\rightarrow\infty$) via the $\zeta$-function renormalization prescription, it is necessary to analytic continue the following function,

\beq
\sum_{i=1}^{L} \Lambda_{i}^{-s} \, {\rm{ln}}\, \Lambda_{i},
\eeq

from large $s$ to $s=0$. Introducing the $\zeta$-function, $\zeta_{P}(s) = \sum_{i=1}^{L} \Lambda_{i}^{-s}$, the logarithm of the determinant of $P$ is completely defined by the first derivative of this $\zeta$-function at $s=0$.
Writing this $\zeta$-function in the usual form of a Mellin transform \cite{SH}\cite{maxent}\cite{zerb},

\beqa
\zeta_{P}(s) &  = & \frac{1}{\Gamma(s)} \int_{0}^{\infty} \, {\rm{Tr}} \, (e^{-JP}) \, J^{s-1} \, {\rm{d}}\!J\\
& \equiv & \int_{0}^{\infty} \, \lambda(\beta,J) \,\, J^{s-1}  \,\, {\rm{d}}\!J
\eeqa
\beq
\label{entropy}
{\rm{ln}}\, {\rm{det}} P  =  -\int^{\infty}_{0} \lambda(\beta,J) \,\, {\rm{ln}}(J) \,\, {\rm{d}}\!J
\eeq

Physically, this expression for the logarithm of the determinant of $P$ in (\ref{entropy}) is related to the entropy of the quantum spin chain, and is the analogue of (\ref{density}) for this quantum system. The logarithm of the determinant of $P$ is equivalently the logarithm of the partition function ${\cal{Z}}$, from (6), and since there is an implicit integration over $\beta$ defined for the continuous-time method integrating over $\beta$ yields a prefactor of $\beta$,

\beq
\int_{0}^{\beta}  {\rm{ln}}\, {\rm{det}} P  \,\,\, {\rm{d}}\beta = \beta  \,\, {\rm{ln}}\, {\rm{det}} P = H
\eeq

where $H$ is of the form of an entropy. However, unlike the classical spin system in (\ref{density}), the total entropy of the quantum spin system is not fully defined from this expression in (\ref{entropy}), and $H$ in (20) only represents the local limit of the entropy of the quantum spin system. In physical terms, this quantum system differs from the classical system in that it is defined via a time-ordered path integral over $\beta$, which in analytical terms, is expressed by the fact that not only is the first moment of (\ref{entropy}) finite, but also a number of the higher moments of the $\zeta$-function are finite. We made an explicit discussion of this problem in \cite{me1}, commenting that the free energy maxima and saddlepoint solutions of a quantum spin system need not be coincident in the general case.

To determine the maxima of the logarithm of the determinant of $P$, and hence the maxima of the support of the system, we have to evaluate the $\zeta$-function prescription at a different point on the fundamental strip, at $s=1$. This is  essentially to include the nonlocal contributions to the path integral within a finite analytic domain of $J$ \cite{maxent}, where,

\beq
\frac{ d^{j} \zeta_{P}(s) }{ ds^{j} } \mid_{s=1}  =  \int_{0}^{\infty} \,\, \lambda(\beta,J) \,\, {\rm {ln}}^{j}(J) \,\,{\rm{d}}\!J, \quad j = 0 ...L
\eeq

The full expression to be analytically continued to determine this support, from (16), is,

\beq
H = \int_{0}^{\infty} \,\, \lambda(\beta,J) \,\, {\rm {ln}}\lambda(\beta,J) \,\,{\rm{d}}\!J
\eeq

which corresponds to the more familiar Von Neumann form of the entropy of a quantum spin system \cite{ww}. This expression is not as analytically well defined as the local expression for the entropy in (\ref{entropy}), because the branch points in the path integral over $\beta$ are unresolved at $s=1$. Hence, in the general case, the Von Neumann entropy can be a divergent function \cite{wil}. However, since we know from the $\zeta$-function prescription in (\ref{entropy}) that it is possible to write down a local limit for the entropy, it is sufficient (to determine the properties of the support that we are interested in) to show that the maximum value of this local limit is unique. In physical terms, this means that we can treat the important case of the quantum system having a degenerate groundstate, by identifying the maxima of the support of the system but without specifying which of these degenerate vacuua we are considering.

In order to establish that both our new definition of $\lambda(\beta,J)$ is a valid FSS function we must, therefore, establish the location of the maxima of the Von Neumann entropy in (22) is asymptotically convergent towards a particular point in $J$ as a function of discrete $J$. To do this we follow the analytic treatment given in \cite{maxent}. The total entropy of the quantum spin system is constructed from the method of Lagrange undetermined multipliers from the (local) moments in (21). Firstly, $\lambda(\beta,J)$ is approximated by a series that contains a finite number of degrees of freedom $M$,

\beq
\lambda(\beta,J) \sim {\rm{exp}}\left( -\sum_{j=0}^{M} \gamma_{j}(\beta) J^{j}\right)
\eeq
Then, the moments of (22) are defined in order to explore the branch structure of $J$,

\beq
\mu_{j} = \int_{0}^{\infty} {\rm{ln}}^{j} \!J \,\, \lambda(\beta,J) \,\, {\rm{d}}J
\eeq

The maxima of (22) is then asymptotically convergent to one fixed value of $J$, as a function of $M$, provided that the following functional is convex,

\beq
\label{gamma}
\Gamma(\gamma_{1}(\beta), ...\, , \gamma_{M}(\beta)) = \sum_{j=0}^{M}
\gamma_{j}(\beta) \, \mu_{j} \nonumber  + \mu_{0} \, {\rm{ln}} \! \left[
\frac{1}{\mu_{0}}
\int_{0}^{\infty} \!\!{\rm{exp}} \! \left( - \sum_{j=0}^{M}
\gamma_{j}(\beta) \, {\rm{ln}}^{j} J \right) {\rm{d}}J \right]
\eeq

What this means is that the Von Neumann entropy in (22) then has a well defined asymptotic limit as a function of any finite local degree of freedom $M$, such as lattice system size, since the normalizing conditions in (14) guarantee the convexity of the above functional \cite{maxent}.

Both $H$ and $\lambda(\beta,J)$ are therefore valid FSS functions for our numerical lattice calculation, and on a practical level, the continuous probability density function $\lambda(\beta,J)$ can be expressed by the number density of the local transition probabilities $\lambda_{i}$. We can therefore identify (19) and (14) as our quantum system analogues of the defining relation for the zeros density of the classical spin system in (\ref{normal}), and its normalizing conditions in (\ref{density}). Our definition for the quantum analogue of the zeros density is the continuous probability density function $\lambda(\beta,J)$. From (19) the zeroes of our new expansion follow exactly the same form of FSS as the zeros of classical systems treated in \cite{z1}-\cite{zn}, and so we can choose a simple difference relation to define $\lambda(\beta,J)$ for our finite lattice systems \cite{lambda3},

\beq
\label{den}
\lambda(\phi) \equiv \lambda \left( \frac{\phi_{k+1}+\phi_{k}}{2} \right) =
\frac{1}{N(\phi_{k+1}-\phi_k)}
\eeq

where $k$ is the sequential index assigned to the zeros along the locus they form in the complex-$(\beta J)^2$ plane, and
$\phi$ is the angle subtended from the real-$(\beta J)^2$ axis to a given zero.
Since $\lambda(\beta,J)$ is a well defined FSS scaling function, the scaling of the zeros density describes the
renormalization group scale transformations of the lattice system,

\beq
\label{asymp1}
\lambda(\beta,J,L) = L^{c} \lambda(\beta L, J L)
\eeq

where $c$ is the scale transformation parameter. As is usual for FSS functions \cite{FSS}\cite{FSS3} in the vicinity of a
second order critical point (\ref{asymp1}), this leads to us approximating that,

\beq
\label{asymp}
\lim_{J,\beta \rightarrow 0} \lambda(J) = J^c ( 1 - a J\, ... \,)
\eeq

where $a$ is an arbitrary constant. As the $\beta$-dependence of $\lambda(\beta, J)$ is continuous, by construction, we can
also make the substitution $e^{i\phi}\equiv J$ to define,

\beq
\label{fit}
{\rm{ln}} \lambda(i\phi) = c \, {\rm{ln}}(J) - aJ  \,\, ...\quad\quad ; \quad\quad
{\rm{ln}} \lambda(i\phi) \sim c / \, \phi
\eeq

Similarly, the first order scaling contribution is of the form,

\beq
\label{spacing}
\Delta\tau_{\,I\!R} = -\frac { {\rm{ln}} \lambda(i\phi)}{\pi} {\big|}_{\lambda (i\phi)\,= \, const.}
\eeq

This form of this function follows from the residue theorem applied to the point where the zeroes density develops its
maxima, which is a nonanalytic point in the infinite lattice volume limit. We label this point, $J_{0}$, and some general
contour around this point in the complex $(\beta J)^{2}$-plane, $\mathcal{C}= e^{i\sigma} + J_{0}$. Therefore,

\beq
\int_{\mathcal{C}} \frac{dJ}{J-J_{0}} = \int_{0}^{4\pi} \frac{ie^{i\sigma} d\sigma}{e^{i\sigma}} = 4\pi i\Delta\tau_{IR},
\eeq

\section{Numerical Results and analysis}

We have generated lattice ensembles the AFM periodic mixed spin chain of the form, $1-1-3/2-3/2$, where $1$ and $3/2$
represent the spin magnitudes and the period of the alternation is four lattice sites, by using the continuous-time method
\cite{me2}\cite{me3}. We consider this model defined with two independent nearest neighbour spin couplings, defined via (1).
We keep the ratio between the like-like and like-dislike neighbouring spin pairs fixed in our analysis and consider the
analytic continuation of the lattice partition function in $\beta$ and $J$. Separate treatments of this model via the
Nonlinear Sigma Model (NLSM) and Density Matrix Renormalization Group (DMRG) methods have indicated that the groundstate of
this system should have three distinct topological regions, with a gapful groundstate, separated by two gapless quantum
critical points at zero temperature \cite{mix1}\cite{mix2} corresponding to the vacuum angles $\theta=\pi$ and $\theta=3\pi$.
In this model $\theta$ can be varied by varying the ratio of the nearest neighbour spin interaction couplings within the
periodic cell. If one of these couplings is fixed to unity the two quantum phase transitions are found at values of the
second coupling of $J_{c}=0.483$ and $J_{c}=1.314$, \cite{mix1}\cite{mix2}. However, neither of these analyses has been able
to quantify the renormalization group flow of the physical couplings to the $\theta=\pi$ IR fixed point.

For our numerical analysis we have generated a range of lattice data sets each consisting of $100,000$ decorrelated lattice
configurations at fixed $\beta$, $J$ and chain length $L$ using the continuous-time method. We have then evaluated the
polynomial expansion coefficients $c_i$ defined in (\ref{poly}) by using the procedure defined in Section 3, using the local
transition probabilities realised in the importance sampling of the continuous-time simulation of the quantum spin chain
system as our numerical input. We have used a standard rootfinder to then calculate the zeros of these polynomials for each
of the lattice ensembles. These zeroes are plotted for two of the ensembles in Fig.1. The general picture is consistent with
our expectation: namely, that the zeroes are constrained to lie on a regular locus in the complex expansion parameter plane,
as we argued in Section 4. The zeroes in Fig.1 are plotted for lattice ensembles generated at two different values of $J$.
The first of these values, $J_{c}=0.483$, corresponds to the region around the quantum phase transition at $\theta=\pi$
whereas the second value, $J_{c}=1.314$, corresponds to the region around the quantum phase transition at $\theta=3\pi$.

\subsection{Critical values of $\beta J$}
We find in Fig.1 that both of the two lattice ensembles not only contain information about the phase transition points
that the lattice ensembles are generated in the vicinity of, but also the second transition that the ensembles are a
considerable distance from in $J$ in the phase space. There is a critical point (indicated by the point at which the
zeroes density (\ref{den}) sharply increases) in both ensembles at both $\phi=0$ and at $\phi=0.9$, where $\phi$ is the angle
subtended from the positive real-axis anticlockwise in the figure. One important feature of our semi-analytic polynomial
expansion is that it enables us to analytic continue the lattice partition function as a function of $\beta J$.
\begin{figure}
\epsfxsize=3.3 in
\centerline{\epsffile{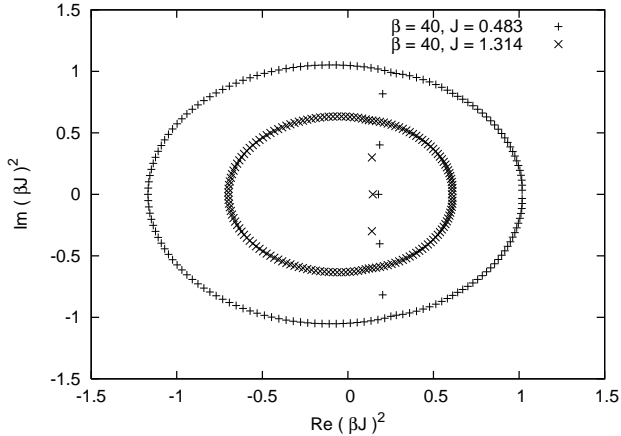}}
\caption{Partition function zeros of the quantum spin chain in the complex $(\beta J)^{2}$ plane, for two different lattice
ensembles. The lattice size for both systems is $L$=64 and the inverse temperature $\beta=40$, whilst one lattice ensemble
is generated at $J=0.483$ and the other at $J=1.314$.}
\end{figure}
Since the zeroes density approaches a maxima at the real axis at, $\phi=0$, the radius of the zeroes locus,
$(\beta J)_{locus}$, is related to the coordinates of the point on the line of constant physics in the $\beta -J$ plane
which passes through the quantum critical point at zero temperature. Naively, we can substitute the fixed value of one of
the lattice couplings used to generate the ensemble (either $\beta$ or $J$) into this value of the radius,
$(\beta J)_{locus}$, in order to define the coordinates of this point in the $\beta - J$ plane. However, practically the
lattice partition function has a finite cutoff in $J$ beyond which the analytic continuation in $J$ is undefined.
We tabulate the radius of the loci in Table 1 for the different lattice ensembles we generated. We find that the radius of the locus is
unity when the ensembles are generated at a $J$ value less than $J_{c}=0.483$, but that the radius of the loci is rescaled
by a specific factor when the ensembles are generated at a $J$ value greater than $J_{c}=0.483$. Furthermore,
the product of this specific factor and the value of $J$ at which the ensembles are generated is unity, as we tabulate in the final
column of Table 1. This indicates that when we generate ensembles at a point in the $\beta - J$ plane where the line of
constant physics that passes through the quantum critical point is above the $J$ cutoff, but, below the $\beta$ cutoff,
then the critical scaling of the partition function is not dimensionful in physical units.

From the final column of Table 1 we find that the lines of constant physics which pass through the two quantum critical
points are located at roughly $J_{c}=0.483(1)$ and $J_{c}=1.314(0)$ in the $\beta -J$ plane. These values can be obtained
independently from all the lattice ensembles we have generated, and are consistent with the critical values found in the DMRG and NLSM calculations in
\cite{mix1}\cite{mix2}.

\begin{table}
\begin{center}
\begin{tabular}{|l|l|l||r|r||}
\hline
$L$        & $\beta$	& $J$      	   & $(\beta J)^{2}_{locus} (\pm 0.001) $ 		&
$J \times (\beta J)_{locus}$ 		\\
\hline
\hline
16	   & 100, 40	& 0.483, 0.483	   & 1.080, 1.080 	& 0.582, 0.582	 		\\
\hline
32	   & 100, 40	& 0.483, 0.483	   & 1.038, 1.038 	& 0.492, 0.492 			\\
\hline
64	   & 100, 40	& 0.483, 0.483	   & 1.018, 1.018 	& 0.487, 0.487	 		\\
\hline
128	   & 100, 40	& 0.483, 0.483	   & 1.004, 1.004 	& 0.484, 0.484 			\\
\hline
\hline
16	   & 100, 40	& 1.314, 1.314	   & 0.702, 0.702 	& 1.101, 1.101	 	 	\\
\hline
32	   & 100, 40	& 1.314, 1.314	   & 0.639, 0.638	& 1.050, 1.050 		 	\\
\hline
64	   & 100, 40	& 1.314, 1.314	   & 0.608, 0.608	& 1.025, 1.025		 	\\
\hline
128	   & 100, 40	& 1.314, 1.314	   & 0.591, 0.591	& 1.010, 1.010		 	\\
\hline
\end{tabular}
\end{center}
\caption{Dependence of the value of the zero nearest the real axis in the complex $(\beta J)^{2}$-plane,
($\beta J)_{locus}$, on the numerical ensemble parameters: lattice size, $L$, inverse temperature $\beta$, and nearest
neighbour coupling $J$. The couplings and results in the columns are matched pairwise across the table. The final column
indicates the relation between the value of the couplings at the critical point, in physical units, and the lattice volume
in physical units.}
\end{table}

\subsection{Critical Scaling Exponents}
We plot the logarithm of our new zeros density function (11) in Fig 2, which we have measured via the simple difference
relation in (15). Two forms of scaling are evident, in line with our expectations. Firstly, in the vicinity of the two
critical points at $\phi=0$ and $\phi=0.9$ the logarithm of the zeros density follows the second order scaling fit of
(\ref{fit}). We tabulate our fitting results to (\ref{fit}) in Table 2. Within the error bars of the analysis, the fitted
value of the exponent is consistent with a value of around one for all of the lattice ensembles. The stability of our
determination of the exponent apparently improves with increasing spatial volume until we reach $L=128$, where the errors
then increase. The error bars on the measurements are in good agreement with similar measurements made from the QMC method
using spin-correlators \cite{sc1}\cite{sc2}\cite{sc3}. The error bars are also in good agreement with recent DMRG and exact
diagonalization studies \cite{st1-dm}\cite{st3-dm}\cite{dm2}. Furthermore, the system size at which our analysis appears to
be come unreliable directly corresponds to the range of volumes that can be treated in these recent studies, which indicates
a potential common dominant source of systematic error in the numerical roundoff errors from diagonalization.

It is interesting to note, however, that the lattice volume dependence of the numerical stability of our measurements
appears to be correlated with the second form of scaling that we have measured in our data, which is of the form of an
overall $\phi$-independent first order contribution to the logarithm of the zeros density, (19). If we compare lattice
ensembles generated at the same fixed $\beta$ and $J$ in Table 2, the scale of the Euclidean-time spacing fluctuation
varies by about 1\% when we go from a spatial volume of $L=64$ to a volume of $L=128$ (in the final column).

The generic origin of this first order contribution to the zeros density is slightly subtle. Naively, the physical volume of
the lattice in the continuous-time method is fixed by choosing the fixed values of $\beta$ and $J$ at which the lattice
ensemble is generated. Generating lattice data sets at different values of $L$, at fixed $\beta$ and $J$, should therefore
correspond to the same physical volume of system. However, because the role of Euclidean-time and the Markov chain are
interchanged via the use of loop-cluster techniques in (\ref{part1}) it is the size of the loop-clusters that are fixed in
physical units and not the Euclidean-time lattice spacing since the Monte Carlo measure is defined in the continuous-time approach by the free energy of the average loop-cluster and not by the free energy of the lattice system \cite{qmc}\cite{math}. 
Our measurements indicate that the expectation of the Euclidean-time lattice
spacing changes discontinuously as we go from spatial volumes of $L=64$ to volumes of $L=128$ in our lattice ensembles. Although the physical lattice volume is fixed, in principle, because the lattice operators have been Wick-rotated the approach is sensitive to changes in the numerically realised IR cutoff of the ensemble in Euclidean-time.

As we discussed in detail in \cite{me}, what is expected for the renormalization group flow of the physical
couplings of the quantum spin chains in the vicinity of the IR fixed point is that there is a Kosterlitz-Thouless transition line in the vicinity of the fixed point at zero temperature. Therefore, potentially what happens for the $L=128$ volumes is that the line of constant physics that is determined from the analysis no longer passes directly through the zero temperature fixed point, for this ensemble, but instead it hits the zero temperature axis some distance from the fixed point. This would change
the values of the fitted exponents for $L=128$. However, whether the change in the numerical stability has a physical origin
or not the additional first order contribution to the ensemble for $L=128$ indicates that there is a change in the condition
number of the system being modeled \cite{ed1}\cite{ed2}\cite{ed3}, and this directly effects the numerical stability of the analysis through roundoff.
\subsection{Numerical Stability}
Importantly, the FSS results obtained in this new analysis have been obtained from a spectral density definition of the
zeroes density function, given in (16), where the scaling is defined as a function of the spectral index, $k$, rather than as
a function of the lattice volume, $L$. This means that we have obtained the critical exponents in Table 2 by using the
information contained within single lattice ensembles generated at just one arbitrary value of the lattice size and couplings. We stress, we
have not combined the information contained in lattice ensembles generated at several different volume sizes to obtain our
exponent results, nor have we combined the information contained in lattice ensembles generated at several different
couplings values, which would be usual for the FSS analysis of Monte Carlo data. We have in fact obtained 24 independent, consistent results for this scaling exponent from 24 independently generated
lattice ensembles of varying lattice sizes and with varying fixed coupling values. We have also performed a parallel QMC analysis
of this model in \cite{me3}, using conventional spin-correlator based measurement techniques, and have found that the computer
time needed to perform this new measurement for one lattice ensemble at a fixed volume and fixed couplings is comparable to
the computer time needed to generate one spin-correlator measurement for one lattice ensemble at a fixed volume and fixed
couplings.

\begin{figure}
\epsfxsize=3.3 in
\centerline{\epsffile{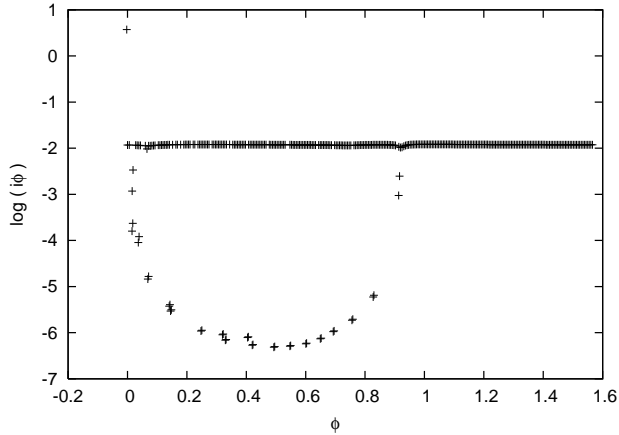}}
\caption{The density of the zeros $\lambda(i\phi)$ in the complex-$(\beta J)^{2}$ plane for an ensemble generated at $L=512$,
$\beta=40$ and $J=0.483$, plotted as a function of the angle, $\phi$, which measures the angle subtended by a zero on the
zeroes locus from the real axis of the complex-$(\beta J)^{2}$ plane (anticlockwise along the ellipse in Figure 1). Two
sorts of scaling are evident in the plot: a constant behaviour corresponding to first order scaling (\ref{spacing}) where the
zeroes density is flat, and a divergent second order scaling behaviour (\ref{fit}) corresponding to the two second order
transition points at $\phi=0$ and $\phi=0.9$ .}

\end{figure}
\begin{table}
\begin{center}
\begin{tabular}{|l|l|l||r|r||r|}
\hline
$L$        & $\beta$	& $J$      & $c (\phi =0)$  & $c (\phi =0.9)$  & $\Delta\tau_{IR}$	\\
\hline
\hline
16	   & 40		& 0.483	   & 1.008(0.078) & 1.008(0.078)  & 0.629(0.217) \\
\hline
16	   & 100	& 0.483	   & 1.008(0.078) & 1.008(0.078)  & 0.628(0.217) \\
\hline
32	   & 40		& 0.483    & 1.002(0.027) & 1.002(0.027)  & 0.627(0.149) \\
\hline
32	   & 40		& 1.314	   & 1.004(0.041) & 1.004(0.041)  & 0.627(0.152) \\
\hline
32	   & 100	& 0.483	   & 1.002(0.027) & 1.002(0.027)  & 0.627(0.149) \\
\hline
32	   & 100	& 1.314	   & 1.011(0.115) & 1.011(0.115)  & 0.625(0.143) \\
\hline
64	   & 40		& 0.483	   & 1.002(0.017) & 1.002(0.017)  & 0.626(0.107) \\
\hline
64	   & 40		& 1.314	   & 1.002(0.004) & 1.002(0.004)  & 0.623(0.109) \\
\hline
64	   & 100	& 0.483	   & 1.002(0.016) & 1.002(0.016)  & 0.626(0.106) \\
\hline
64	   & 100	& 1.314	   & 1.002(0.009) & 1.002(0.009)  & 0.623(0.109) \\
\hline
128	   & 40		& 0.483	   & 1.216(0.218) & 1.216(0.218)  & 0.616(0.088) \\
\hline
128	   & 100	& 0.483	   & 1.163(0.193) & 1.163(0.193)  & 0.615(0.077) \\
\hline
\hline
\end{tabular}
\end{center}
\caption{Dependence of the second order scaling exponent $c$ on the numerical simulation parameters: lattice system size
$L$, inverse temperature $\beta$, and nearest neighbour coupling $J$. The exponent $c$ is determined from a fit of the
logarithm of the zeros density ${\rm{ln}}\lambda(i\phi)$ defined in (\ref{den}) to the second order scaling function defined
in (\ref{fit}). The value of $\Delta\tau_{IR}$, which measures the scale of the nonlocal fluctuations of the lattice
ensemble, is determined from a fit to (\ref{spacing}).}
\end{table}

\section{Nonperturbative Renormalization}

Importantly, our new analysis defines the analytic continuation of the lattice partition of the continuous-time QMC method
as a function of $\beta$ and $J$. This then implies that the second order scaling behaviour that we have identified from the
partition functions of our analysis corresponds to the zero temperature IR point of the quantum spin chain system at
$\theta=\pi$, since we have Wick-rotated the spin operator definition via the exact mapping $\beta \rightarrow i\beta$. Although we have generated our numerical data at finite temperature, via this new analysis, it
is therefore possible for us to investigate the zero temperature IR point of the quantum spin chain system at $\theta=\pi$
directly using finite-temperature lattice ensembles. We are able to do this, essentially, by defining the line of constant
physics which passes through the point defined by the lattice parameter values in the $\beta - J$ plane and the IR fixed
point.

In \cite{me} and \cite{me7} we have recently treated the renormalization of the generic quantum spin chain system in the
vicinity of this IR fixed point at $\theta=\pi$, using a lattice perturbation theory formalism. Recent
conformal arguments have suggested that in addition to the usual local Sine-Gordon model effective field theory picture of
the quantum spin chain in this region \cite{afm}\cite{fat}, it is also important to further consider nonlocal quantum
effects in this region by generalising these effective field theory treatments to the double Sine-Gordon model \cite{mus}. Our
analysis in \cite{me} and \cite{me7} suggests that, for quantum fluctuations of arbitrary size, the double Sine-Gordon model
should generically require nonperturbative renormalization. The analytic continuation properties of the lattice partition we
have defined in this article essentially define that nonperturbative renormalization program. We quantify the simple local
fluctuations of the quantum spin chain via the second order scaling contribution, but in addition, we also quantify the
relative size of the nonlocal quantum fluctuations via the nonanalytic contribution $\Delta\tau_{IR}$. Therefore, we can use
this scheme to relate any arbitrary point in the $\beta-J$ coupling plane of the experimental system that this lattice model
represents to the critical scaling at $\theta=\pi$, and quantify the scale of the quantum fluctuations in physical units.

\section{Entanglement entropy}

As we discussed in Section 5, the defining relation for the zeroes density  of the lattice partition function of the quantum spin chain system in (\ref{entropy}) is an expression for an entropy. Specifically, it is the regularized local limit of the more usual Von Neumann entropy in (22), defined via a $\zeta$-function renormalization prescription \cite{SH}\cite{maxent}\cite{zerb}. To obtain (\ref{entropy}) we have considered the continuous-$J$ limit of the lattice partition function in (6), in which the local transition probabilities $\lambda_{i}$ become a continuous function of $J$, given by $\lambda(\beta,J)$. We have then shown explicitly that this continuous-limit can be reached from a discrete lattice system. 
Importantly, this definition connects with how the Von Neumann entropy in \cite{wil}\cite{ww} describes how the subsystem of a quantum spin system is entangled with the remainder of the system. From \cite{wil}, these subsystems are defined for our system via the intervals $[0,J[$ and $[J,\beta[$, where the full quantum system (universe) is defined via the region $[0,\beta[$. From the meromorphic continuation properties rigorously justified in Section VB, it follows that the entanglement entropy of the discrete quantum spin chain lattice system is given by,

\beq
\label{jam}
H  =  -{\rm{Tr}} \left[ \, \lambda(\beta,J) \,\, {\rm{ln}} \lambda(\beta,J) \,\, \right]
\eeq

where the entanglement is defined between the two subsystems of the lattice system defined by the bipartite Trotterization of the lattice in (3).

Several conformal field theory results have been derived for the FSS of the entanglement entropy of quantum spin chains \cite{wil}\cite{card}. The basic result is that the entanglement entropy of a quantum spin chain scales like the logarithm of the physical area of the system. This area can be parameterised by the physical lattice units of the partition function in (\ref{part1}), and the second order scaling result we have found is that the scaling of the zeroes density (which is directly related to the entanglement entropy via (22)) is directly proportional to the area of the lattice in physical units of $J$, see Table 1. However, the strength of our analysis is that we are able to further quantify the nonlocal corrections to this simple scaling pattern in the vicinity of the IR fixed point of the quantum spin chain system at $\theta=\pi$, and we are able to use then use this scaling result to give precise estimates for the central charge of the lattice ensembles we have generated numerically.

It is known from conformal field theory arguments that the central charge, $c$, of the general quantum spin chain system acquires a very specific value at the point $\theta=\pi$, and that this point is related to the unstable renormalization group flow of the SU(2) Wess-Zumino-Witten model and the emergence of a gapless groundstate in half-integer spin chains. According to Haldane and Affleck \cite{afm}, gapped groundstates states are realised by quantum spins chains with integer spin values. However, spins chains with half-integer spin values are all gapless, and all correspond to the same fixed point of the SU(2) Wess-Zumino-Witten model with the same central charge value of $c=1$. In the mixed-spin quantum spin model we have treated it is known that varying the ratio of the nearest-neighbour spin interaction coupling between the spins of different magnitudes ($\alpha$ in (1)) drives the model through the vacuum angle $\theta=\pi$. This is evidenced by the fact that the finite gap becomes vanishing at specific values of this coupling ratio \cite{me2}\cite{me3}\cite{mix1}\cite{mix2}. Varying the ratio couplings in this model, therefore, interpolates between these two different physical states of the generalised quantum spin chain system: gapped and gapless. Strictly, however, no measurement of the inverse spin gap at this gapless point can be truly a divergent function in a finite-temperature lattice data set as a function of increasing lattice volume, and this measurement is only strictly a divergent function in these calculations upto the intrinsic IR cutoff scale of the finite system. This explicit cutoff scale dependence is quantified in our measurements via $\Delta\tau_{IR}$. This nonanalytic term and correction to the area law FSS of the entanglement entropy of quantum spin chains has been also identified in Density Matrix Renormalization Group and Exact Diagonalization calculations in \cite{st1-dm}\cite{st2}\cite{st3-dm}.

The specific value of the scaling exponent of the entanglement entropy of quantum spin chains is $\chi (c + \overline{c})/6$ from conformal field theory calculations \cite{wil}\cite{card}\cite{euler}, where $\chi$ is a
geometric factor known as the Euler number which parameterises the topology of the boundary between spins, and
$\overline{c}$ is antiholomorphic central charge of the quantum spin chain system. In our calculation $\overline{c}=0$, and $\chi=6$. This first value follows because the boundary conditions for Euclidean-time are defined as $(0,\beta]$, and so only forward propagating modes along the Markov chain are defined analytically via (19). This corresponds to the physical problem discussed in \cite{wil}, of the entropy of a conformal system defined by reflections to the right of a moving mirror. The second value, for the Euler number, follows from our discussion in Section 2. The basic form of the continuous-time method is unaffected by including spin vectors of larger magnitude, which can be simply incorporated by a permutation factor \cite{lu}. Therefore, even though we have considered a nonuniform quantum spin chain for our analysis with periodic cells of the form $1-1-3/2-3/2$, the geometry of the boundaries between two subsystems is unaffected by this choice of mixed-spin model. Nearest-neighbour spin sites generically give an Euler number of two \cite{euler} and so for the lattice partition function of the quantum spin chain defined in (\ref{matrix}) there are essentially three independent nearest-neighbour sites, giving a generic $\chi=6$ for this method. Therefore, our conjecture that the scaling exponent we have measured describes the second order scaling of the quantum spin chain system in the vicinity of the point $\theta=\pi$ seems to be born out by the numerics. The prediction of Affleck and Haldane for the central charge of this IR fixed point in the generalised quantum spin chain system is $c=1$, and we have measured $c=1$ to within 20\% for all the lattice ensembles we have generated. Discounting the larger volumes at $L=128$ this result is further improved to agreement with unity to within 1\% for all data sets. The size of the error bars on these measurements are also in good agreement with the results of recent DMRG and exact diagonalization studies.

\section{summary}
We have presented a new technique to identify the partition function zeros of quantum critical points directly from the
information contained in the numerical transfer matrix of the continuous-time Quantum Monte Carlo method. We have applied
this scheme to a mixed spin quantum spin chain model with a nearest neighbour AFM coupling of the form $1-1-3/2-3/2$, and
have used the scheme to quantify the renormalization group flow of the physical lattice couplings $\beta$ and $J$ of the
model in the vicinity of the $\theta=\pi$ IR fixed point of the quantum spin chain system. Our new technique separates the
local and nonlocal degrees of freedom of a nonperturbative lattice system analytically, and this has allowed us to
investigate the renormalizability of the recent proposal that the $\theta=\pi$ fixed point of quantum spin chains is best
described by the double Sine-Gordon model \cite{mus}. In Section 3 we have given a generic treatment of the continuous-time
method lattice partition function so that our new expansion can readily be generalised to other low-dimensional quantum spin
systems. In Section 7 we have argued that (\ref{entropy}) defines the local limit of the entanglement entropy of our quantum spin chain system defined via a bipartite Trotterization
and, therefore, that we can compare our numerical results with conformal predictions for the FSS of the entanglement entropy.
We have then done this, and have found good agreement between our measured scaling exponents and those predicted by conformal
field theory. We have identified the correct value of the central charge of $c=1$ for the local part of the effective field
theory description of the quantum spin chain in the vicinity of the $\theta=\pi$ fixed point of the system predicted by
conformal field theory \cite{card}\cite{wil}, but we have also rigorously quantified the nonanalytic correction due to nonlocal
fluctuations. Practically this result can be used to relate the experimental couplings of a quantum spin chain to the
underlying conformal symmetries of the system via a program of nonperturbative renormalization.

\end{document}